\documentclass[%
 superscriptaddress,
 prl,
 amsmath,amssymb,
reprint,%
 author-year,%
longbibliography
]{revtex4-2}

\usepackage{graphicx,dcolumn,amsmath,amssymb,color,mathrsfs,tabularx,titlesec,hyperref,enumerate,bm,subfigure,float,comment,soul}
\definecolor{linkColor}{rgb}{0.7,0,0}
\definecolor{darkred}{rgb}{0.7,0,0}
\hypersetup{pdfborder={0 0 0},colorlinks=true,urlcolor=linkColor,citecolor=linkColor}

\newcommand{\ii}{{i\mkern1mu}}

\newcommand{\angstroms}{\text{\normalfont\AA}}
\newcommand{\f}[2][]{\mathcal{F}_{#1}\left[#2\right]}
\renewcommand{\vec}[1]{\mathbf{#1}}
\newcommand{\mvec}[1]{\bm{#1}}
\newcommand{\smatrix}[0]{$\mathcal{S}$-matrix}
\newcommand{\finv}[2][]{\mathcal{F}_{#1}^{\dagger} \left[#2\right]}

\titleformat{\section}
{\color{darkred}\sffamily\bfseries}
{\color{darkred}\thesection}{1em}{}
\titleformat{\subsection}
{\color{darkred}\sffamily\itshape}
{\color{darkred}\thesection}{1em}{}

\emergencystretch=\maxdimen
\usepackage{algorithm,algorithmic}
\usepackage{mathtools}
\usepackage{physics}

\usepackage{siunitx}

\begin{document}


\title[PRISM 2.0]{A Fast Algorithm for Scanning Transmission Electron Microscopy (STEM) Imaging and 4D-STEM Diffraction Simulations}

\author{Philipp M. Pelz}
\email{philipp.pelz@berkeley.edu }
\affiliation{Department of Materials Science and Engineering, University of California Berkeley, Berkeley, CA 94720}
\affiliation{National Center for Electron Microscopy, Molecular Foundry, Lawrence Berkeley National Laboratory, 1 Cyclotron Road, Berkeley, CA, USA, 94720}

\author{Alexander Rakowski}
\affiliation{National Center for Electron Microscopy, Molecular Foundry, Lawrence Berkeley National Laboratory, 1 Cyclotron Road, Berkeley, CA, USA, 94720}

\author{Luis Rangel DaCosta}
\affiliation{Department of Materials Science and Engineering, University of California Berkeley, Berkeley, CA 94720}
\affiliation{National Center for Electron Microscopy, Molecular Foundry, Lawrence Berkeley National Laboratory, 1 Cyclotron Road, Berkeley, CA, USA, 94720}

\author{Benjamin H Savitzky}
\affiliation{National Center for Electron Microscopy, Molecular Foundry, Lawrence Berkeley National Laboratory, 1 Cyclotron Road, Berkeley, CA, USA, 94720}

\author{Mary C Scott}
\affiliation{Department of Materials Science and Engineering, University of California Berkeley, Berkeley, CA 94720}
\affiliation{National Center for Electron Microscopy, Molecular Foundry, Lawrence Berkeley National Laboratory, 1 Cyclotron Road, Berkeley, CA, USA, 94720}

\author{Colin Ophus}
\email{cophus@gmail.com}
\affiliation{National Center for Electron Microscopy, Molecular Foundry, Lawrence Berkeley National Laboratory, 1 Cyclotron Road, Berkeley, CA, USA, 94720}

\date{\today}

\begin{abstract}

Scanning transmission electron microscopy (STEM) is an extremely versatile method for studying materials on the atomic scale. Many STEM experiments are supported or validated with electron scattering simulations. However, using the conventional multislice algorithm to perform these simulations can require extremely large calculation times, particularly for experiments with millions of probe positions as each probe position must be simulated independently. Recently, the PRISM algorithm was developed to reduce calculation times for large STEM simulations. Here, we introduce a new method for STEM simulation: partitioning of the STEM probe into ``beamlets,'' given by a natural neighbor interpolation of the parent beams. This idea is compatible with PRISM simulations and can lead to even larger improvements in simulation time, as well requiring significantly less computer RAM. We have performed various simulations to demonstrate the advantages and disadvantages of partitioned PRISM STEM simulations. We find that this new algorithm is particularly useful for 4D-STEM simulations of large fields of view. We also provide a reference implementation of the multislice, PRISM and partitioned PRISM algorithms.

\end{abstract}

\pacs{PACS Numbers}
\maketitle

\section{Introduction}
\label{sec:intro}

Transmission electron microscopy is a powerful tool for studying atomic-scale phenomena due to its unmatched spatial resolution, and ability to perform imaging, diffraction, and multiple types of spectroscopic measurements \cite{egerton2005physical, carter2016transmission, zuo2017advanced}. Scanning TEM (STEM) is a particularly versatile TEM technique, as the STEM probe size can be tuned to any desired experimental length scale, from sub-{\AA}ngstrom to tens of nanometers, to best match the length scale of the structures being probed \cite{pennycook2011scanning}. The size of the probe is also completely decoupled from the step size between adjacent probe positions, allowing experimental fields-of-view up to almost one square millimeter \cite{kuipers2016large}. Advances in detector technology have lead to high speed electron cameras capable of recording full 2D images of the diffracted STEM probe with microsecond-scale dwell times, which has lead to many experiments which record the full four-dimensional dataset, in a family of methods called 4D-STEM \cite{ophus2019four}. In parallel, the rise of powerful computational methods have enabled measurements of many different material properties with high statistical significance \cite{spurgeon2020towards}. 

The combination of computational methods and advanced STEM experimentation has lead to atomic-resolution 3D tomographic reconstructions \cite{yang2017deciphering}, measurements of highly beam sensitive samples over functional length scales \cite{panova2019diffraction},  images of samples with resolution better than the diffraction limit \cite{chen2021electron}, and many other advances in STEM imaging techniques. Many of the technique developments and validation of these experiments make heavy use of electron scattering simulations. The application of extremely data-intensive machine learning methods which to STEM experiments can also be aided by simulations \cite{kalinin2021review}. 

It is possible to simulate the propagation and scattering of STEM probes through a material by directly computing the Bloch wave eigenstates of the electron scattering matrix (\smatrix{}) \cite{bethe1928theorie}. The Bloch Wave method can be employed in diffraction simulations \cite{zuo2021cloudemaps}, but it is only practical to use for small, periodic unit cells. The majority of the STEM simulations performed currently implement the multislice method \cite{cowley1957scattering}. The multislice method is typically applied in STEM simulations by performing a separate quantum-mechanical electron scattering simulation for each probe position \cite{croitoru2006efficient, okunishi2012experimental, ophus2016efficient}. The multislice algorithm can therefore require extremely large computation times when simulating STEM experiments which can contain $1000^2$ probe positions or even higher. To alleviate this issue, various authors have implemented parallelized simulation codes that make use of multiple central processing unit (CPU) or graphics processing unit (GPU) resources \cite{grillo2013stemcell, van2015fdes,  hosokawa2015benchmark, allen2015modelling, lobato2016progress, kirkland2016computation, oelerich2017stemsalabim, barthel2018dr, radek2018stemcl}.

It is possible to perform large STEM simulations more efficiently by computing them as a superposition of plane waves \cite{chen1995modification}. This idea was developed into an efficient simulation algorithm by \cite{ophus2017fast}, who named it the plane-wave reciprocal-space interpolated scattering matrix (PRISM) algorithm. In the PRISM algorithm, the \smatrix{} elements are directly computed by multislice simulations. The equivalence of the Bloch wave \smatrix{} and multislice simulation outputs have been investigated in detail by \cite{allen2003lattice, findlay2003lattice}. The PRISM algorithm has been implemented into multiple simulation codes \cite{pryor2017streaming, madsen2020abtem, brown2020python}. It has also been extended to a double-\smatrix{} formalism which can provide an even higher speed boost relative to multislice for inelastic  scattering such as in STEM electron energy loss spectroscopy (STEM-EELS) simulations \cite{brown2019linear}.

The PRISM algorithm achieves large decreases in calculation times by reducing the sampling of the probe wavefunction in reciprocal space, and is  highly accurate when the detector configuration is given by large monolithic regions. However, PRISM simulations are less accurate where fine details in the STEM probe and diffracted Bragg disks are necessary, for example in \cite{juchtmans2015using, hubert2019structure,  zeltmann2020patterned}. A different form of interpolation has been proposed by \cite{pelz2020phase}, where the STEM probe is partitioned into different beams by interpolation of basis functions constructed from the initial STEM probe. This partitioning of the probe has been shown to be a highly efficient and accurate representation of dynamical scattering of the STEM probe in experimental data, and is fully compatible with the PRISM algorithm \cite{pelz2020reconstructing}.

In this manuscript, we introduce the partitioned PRISM algorithm for use in STEM simulations. We describe the theory of multislice, PRISM, and partitioned PRISM simulations, and provide a reference implementation of these algorithms. We show that beam partitioning simulations provide an excellent trade off between calculation times and accuracy, by measuring the error of diffracted STEM probes with respect to multislice simulations as a function of the number of included beams. We also use this method to simulate the full field of view for a common experimental geometry, a metal nanoparticle resting on an amorphous substrate. These simulations demonstrate that the partitioned PRISM method can produce comparable accuracy for coherent diffraction to PRISM simulations, but for much lower calculation times and lower random access memory (RAM) usage. This is important since many PRISM simulations are constrained by the available RAM of a GPU to hold the \smatrix{}. Finally, we demonstrate the utility of this method in 4D-STEM simulations by simulating the full 4D dataset of an extremely large (512$^2$ probes, 4.6 million atoms) sample cell and measuring the sample strain, where the partitioned PRISM algorithm provides superior performance to a PRISM simulation using roughly the same total calculation time.


\section{Theory}

For previously published TEM simulation methods, we will  briefly outline the required steps here. We refer readers to Kirkland for more information on these methods \cite{kirkland2020}. We will also only describe the scattering of the electron beam while passing through a sample; probe-forming optics and the microscope transfer function mathematics are described in many other works \cite{kirkland2020,Williams_Carter_2009,Spence_2013}.

\subsection{Elastic Scattering of Fast Electrons}

Transmission electron microscopy simulations aim to describe how an electron wavefunction $\psi(\vec{r})$ evolves over the 3D coordinates $\vec{r} = (x,y,z)$.  The evolution of the slow-moving portion of the wavefunction along the optical axis $z$ can be described by the Schr\"{o}dinger equation for fast electrons \cite{kirkland2020}
\begin{equation}
    \frac{\partial }{\partial z}\psi(\vec{r}) = 
    \frac{\ii \lambda}{4 \pi} {\nabla_{xy}}^2 \psi(\vec{r})
    + \ii \sigma V(\vec{r}) \psi(\vec{r}),
    \label{eq:paraxial_schroedinger}
\end{equation}
where $\lambda$ is the relativistic electron wavelength, ${\nabla_{xy}}^2$ is the 2D Laplacian operator, $\sigma$ is the relativistic beam-sample interaction constant and $V(\vec{r})$ is the electrostatic potential of the sample.

\subsection{The Bloch Wave Algorithm}

The Bloch wave method uses a basis set that satisfies Eq.~\ref{eq:paraxial_schroedinger} everywhere inside the sample boundary, which is assumed to be periodic in all directions. This basis set is calculated by calculating the eigendecomposition of a set of linear equations that approximate Eq.~\ref{eq:paraxial_schroedinger} up to some maximum scattering vector $|q_{\rm{max}}|$. Then, for each required initial condition (such as different STEM probe positions on the sample surface), we compute the weighting coefficients for each element of the Bloch wave basis. Finally, the exit wave after interaction of the sample is calculated by multiplying these coefficients by the basis set. This procedure can be written in terms of a scattering matrix (\smatrix{}) as \cite{kirkland2020}

\begin{equation}
    \psi_f(\vec{r}) = \mathcal{S} \; \psi_0(\vec{r}),
    \label{eq:Bloch}
\end{equation}
where $\psi_0(\vec{r})$ and $\psi_f(\vec{r})$ are the incident and exit wavefunctions respectively. The Bloch wave method can be extremely efficient for small simulation cells, such as periodically tiled crystalline materials. High symmetry is also an asset for Bloch wave simulations, as it allows the number of beam plane waves (beams) included in the basis set to be limited to a small number. However, for a large STEM simulation consisting of thousands or even millions of atoms in the simulation, the \smatrix{} may contain billions or more entries, which requires an impractical amount of time to calculate the eigendecomposition (roughly $\Theta(B^3)$ for $B$ beams). Using Eq.~\ref{eq:Bloch} many times for multiple electron probes could require extremely large computational times. Thus Bloch wave methods are only used for plane wave, diffraction, or very small size STEM simulations.



\subsection{The Multislice Algorithm}


The formal operator solution to Eq.~\ref{eq:paraxial_schroedinger} is given by \cite{kirkland2020}, 
\begin{equation}
    \psi_f(\vec{r}) = 
    \exp \left\{
        \int_{0}^{z} \left[
        A(z') + B(z') 
        \right] dz'
    \right\}
    \psi_0(\vec{r}), \nonumber
    \label{eq:formal_soln}
\end{equation}
where $\psi_f(\vec{r})$ is the exit wavefunction after traveling a distance $z$ from the initial wave $\psi_0(\vec{r})$. This expression is commonly approximately solved with the multislice algorithm first given by Cowley and Moodie \cite{cowley1957scattering}, which alternates solving the two operators using only the linear term in the series expansion of the exponential operator.

\begin{figure}[htbp]
    \centering
    \includegraphics[width=3.4in]{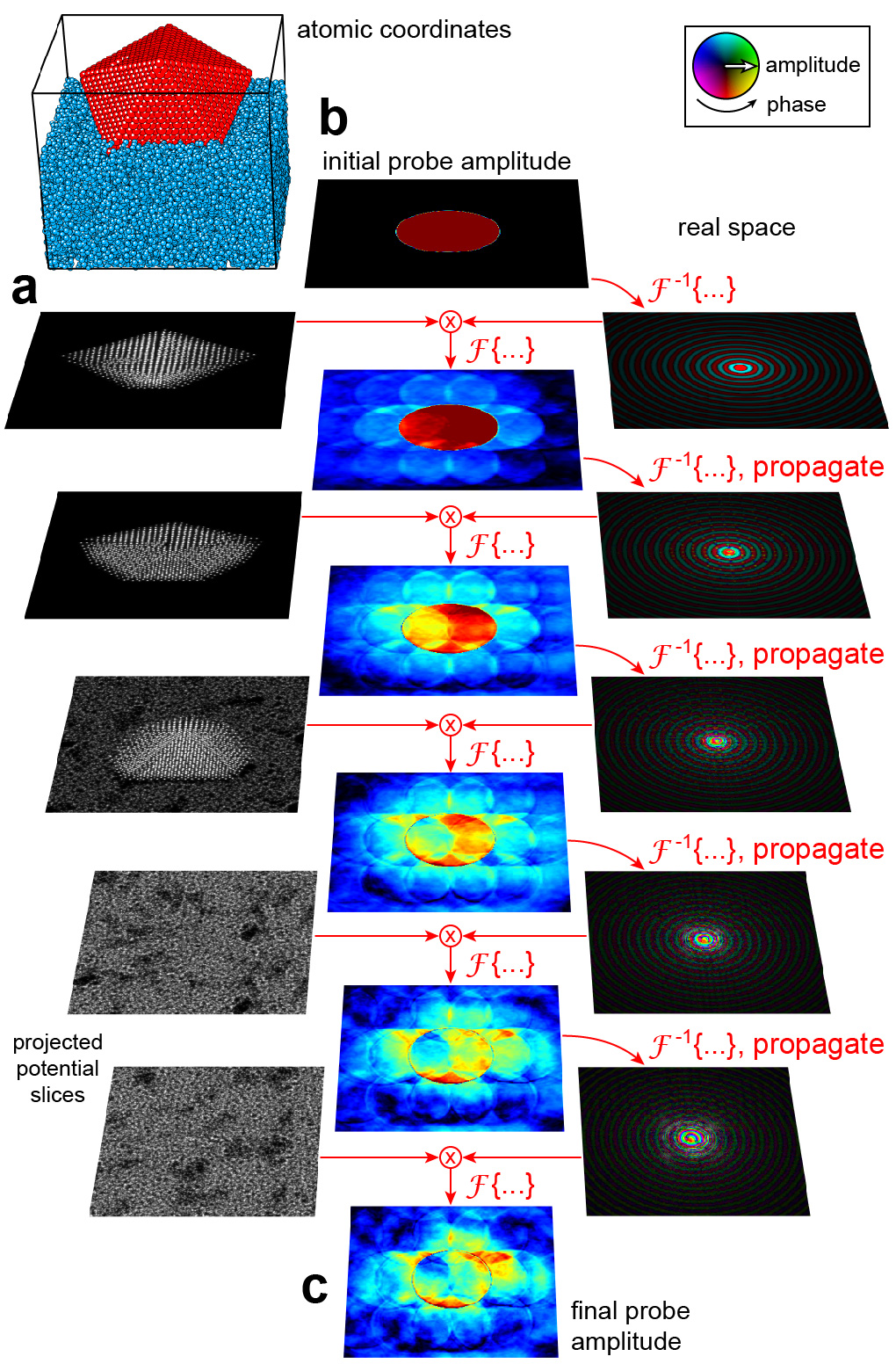}
    \caption{{\bf The multislice simulation algorithm.} (a) Calculate the projected potential slices from the atomic coordinates and lookup tables. (b) Initialize the probe wavefunction, and then alternate between propagation and transmission operators. (c) Final probe at sample exit plane.}
    \label{Figure:multislice}
\end{figure}

In the multislice algorithm, we first divide up the sample into a series of thin slices with thickness $t$. Solving for the first operator on Eq.~\ref{eq:formal_soln} yields an expression for free space propagation between slices separated by $t$, with the solution given by 
\begin{equation}
    \psi_f(\vec{r}) = \mathcal{P}^t \psi_0(\vec{r}),
\end{equation}
where $\mathcal{P}^t$ is the Fresnel propagator defined by
\begin{equation}
    \mathcal{P}^t\psi:=
    \finv[\vec{q}]{\f[\vec{r}]{\psi}e^{-i \pi \lambda \vec{q}^2 t}},
\end{equation}
where $\vec{q}=(q_x,q_y)$ are the 2D Fourier coordinates and $\vec{r}=(x,y)$ are the 2D real space coordinates.
$\f[\vec{x}]{\,\cdot\,}$ denotes the two-dimensional Fourier transform with respect to $\vec{x}$ and $\finv[\vec{x}]{\,\cdot\,}$ the 2D inverse Fourier transform with respect to $\vec{x}$.

To solve for the second operator in Eq.~\ref{eq:formal_soln}, integrate the electrostatic potential of the sample over the slice of thickness $t$
\begin{equation}
    V_{t}(x,y,z) = \int_{z}^{z+t}V(x,y,z')dz'.
\end{equation}
Fig.~\ref{Figure:multislice}a shows an example of this slicing procedure. If we assume that the electron scattering inside this slice occurs over infinitesimal thickness, the resulting solution to this operator is
\begin{equation}
    \psi_f(\vec{r}) = \exp[
    \ii \sigma V_{t}(\vec{r})
    ] \psi_0(\vec{r}).
\end{equation}

We can then write one iteration of the multislice algorithm as 
\begin{equation}
        \label{eq:multiply_propagate}
      \mathcal{T}(\psi,V_{k}) = \mathcal{P}^t\left[\psi\cdot e^{i\sigma V_{k}}\right] := \mathcal{T}_k \psi
\end{equation}
where $V_{k}$ is the projected potential at slice $k$. This algorithm is shown schematically in Fig.~\ref{Figure:multislice}b. The multislice solution of a wavefunction $\psi$ after $k$ potential slices is then 
\begin{equation}
    \mathcal{M}^{V}_k \psi= 
\begin{cases}
    \psi\cdot e^{\ii \sigma V_{k}},& \text{if } k=0\\
    \mathcal{T}(\mathcal{M}_{k-1}(\psi,V),V_{k}),              & \text{if } k>0
\end{cases}
\end{equation}
for which we introduce the short notation $\mathcal{M}_k$ if the potential is assumed to be fixed. It is important to note that $\mathcal{T}(\psi,V_{k})$ is linear in $\psi$ and nonlinear in ${V}_k$ and thus $\mathcal{M}^{V}_k$ is linear in $\psi$ and nonlinear in $V$. Traditionally, a STEM or 4D-STEM simulation was computed by shifting the incoming wave function to the scan positions $\mvec{\rho}$ and computing the resulting far-field intensity using the multislice algorithm at each position:
\begin{equation}
\label{equ:scanning_intensity}
    I(\vec{q},\mvec{\rho}) =  \left|\f[\vec{r}]{\mathcal{M}_k\psi(\vec{r}-\mvec{\rho})}\right|^2.
\end{equation}
An example of this output is shown in Fig.~\ref{Figure:multislice}c. This requires us to perform a full multislice calculation at each scan position, which makes large fields of view that could contain millions of probe positions computationally expensive.

\subsection{The PRISM Algorithm for STEM Simulations}

Recently, \cite{ophus2017fast} proposed an elegant solution to this problem. The incident wave-function of a microscope in a scanning geometry usually passes through a beam-forming aperture with maximum allowed wave vector $h_{max}$ and is then focused onto the sample. It can therefore be described in Fourier space as 
\begin{equation}
\label{equ:fourier_shift_theorem}
    \ket{\psi}_{\vec{r}-\mvec{\rho}} = \sum_{|\vec{h}|<h_{\text{max}}}\Psi(\vec{h})e^{2 \pi \ii \vec{h}\cdot(\vec{r}-\mvec{\rho})},
\end{equation}
with $\Psi(\vec{h})$ the Fourier transform of $\psi(\vec{r})$ and $\mvec{\rho}$ the two-dimensional scan coordinate.  Using the linearity of the multislice algorithm with respect to $\psi$ and Eq.~\ref{equ:fourier_shift_theorem}, we can then rewrite Eq.~\ref{equ:scanning_intensity} as 
\begin{equation}
    \label{equ:phase_shifting_multislice1}
    I(\vec{q},\mvec{\rho}) = \left|\f[\vec{r}]{\sum_{\vec{h}<h_{max}}\Psi(\vec{h})e^{-2\pi i\vec{h}\cdot\mvec{\rho}}\mathcal{M}_k e^{2 \pi i\vec{h} \cdot\vec{r}}}\right|^2.
\end{equation}
We take Eq.~\ref{equ:phase_shifting_multislice1} to link our algorithm to the existing Bloch wave literature in electron microscopy. Traditionally, the set $\vec{h}$ of incoming plane waves is referred to as ``beams,'' and the linear operator that maps from plane waves entering he sample to plane waves exiting the sample is referred to as the \smatrix{}. Using Eq.~\ref{equ:phase_shifting_multislice1}, we can define the real-space scattering matrix ${\mathcal{S_{\vec{r},\vec{h}}} := \mathcal{M}_k e^{2 \pi i\vec{h}\cdot\vec{r}}}$, which is the set of exit waves produced by running the multislice algorithm on the set of plane waves present in the probe-forming aperture of the microscope. The scattering matrix encapsulates all amplitude and phase information that is required to describe a scattering experiment with variable illumination, given a fixed sample potential $V$.

Given the \smatrix{} and a maximum scattering angle $h_{\text{max}}$ in the condenser aperture, we can rewrite Eq.~\ref{equ:phase_shifting_multislice1} with the real-space scattering matrix as 
\begin{equation}
    \label{equ:phase_shifting_multislice2}
    I(\vec{q},\mvec{\rho}) = \left|\f[\vec{r}]{\sum_{\vec{h}<h_{max}}\Psi(\vec{h})e^{-2 \pi i\vec{h}\cdot\mvec{\rho}}\mathcal{S}_{\vec{r},\vec{h}}}\right|^2.
\end{equation}
To introduce the concepts used in the PRISM algorithm we now need to consider the variables $\vec{r}$ and $\vec{h}$ on a discretely sampled grid. The bandwidth-limitation $|\vec{h}|<h_{\text{max}}$ means that the incoming probe is represented by a finite number of Fourier coefficients ${\vec{h}_{\mathsf{b}} \in \mathsf{H} = \{(h_x,h_y)\,|\,||\vec{h}||_2<h_{\text{max}}}\}$. Let the discretely sampled \smatrix{} have dimensions $\mathcal{S}_{\vec{r},\mathsf{b}} \in \mathbb{C}^{\mathsf{N}_1 \cross \mathsf{N}_2 \cross \mathsf{B}}$, with $\mathsf{N}_1 \cross \mathsf{N}_2$ the real-space dimensions and $\mathsf{B} = |\mathsf{H}|$ the number of pixels sampled in the condenser aperture.

\begin{figure}[htbp]
    \centering
    \includegraphics[width=3.4in]{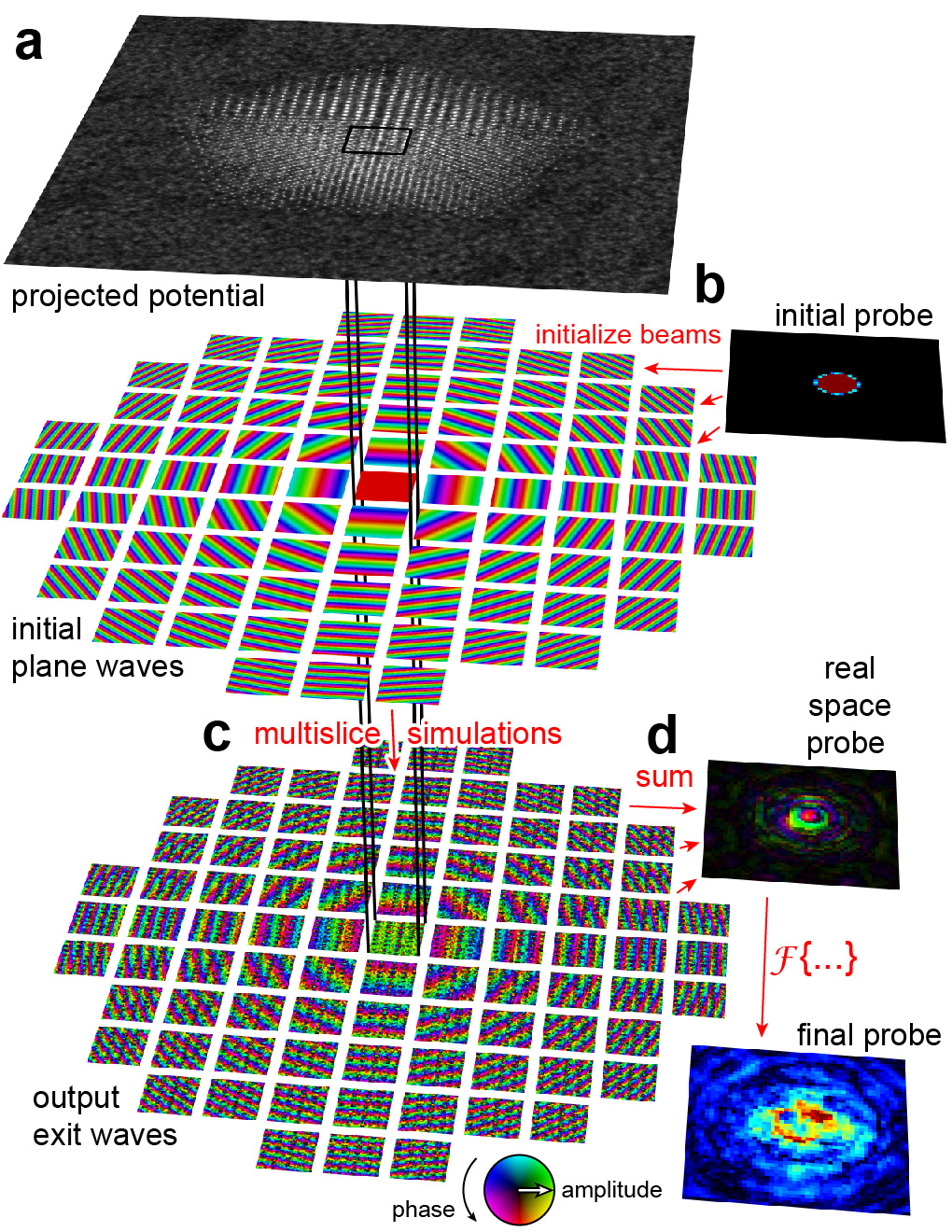}
    \caption{{\bf The PRISM simulation algorithm.} (a) Calculate the projected potential slices from the atomic coordinates and lookup tables. (b) Select an interpolation factor $f$ and a maximum scattering angle $|q_{\rm{max}}|$, initialize all tilted plane waves needed for these beams. (c) Perform a multislice simulation for each beam over the full field of view, store in the \smatrix{}. (d) Compute outputs by shifting the initial STEM probes and cropping $1/f$ of the total field of view, and multiplying and summing all \smatrix{} beams.}
    \label{Figure:PRISM}
\end{figure}

To compute the \smatrix{}, we run the multislice algorithm for each wavevector $\vec{h}_{\mathsf{b}}$ that is sampled in the detector plane. These steps are shown schematically in Figs.~\ref{Figure:PRISM}a and b. This strategy yields favorable computational complexity when a large number of probe positions needs to be calculated, which is necessary for large field-of-view STEM simulations. It has the additional advantage that a series of scanning diffraction experiments with different illumination conditions can be simulated without recomputing the \smatrix{}. This method was named the plane-wave reciprocal-space interpolated scattering matrix (PRISM) algorithm in \cite{ophus2017fast}. It was first implemented into a simulation code  parallelized for both CPUs and GPUs in the Prismatic implementation \cite{pryor2017streaming}. Since then, the PRISM algorithm has also been implemented in the GPU simulations codes py\_multislice \cite{brown2020python} and abTEM \cite{madsen2020abtem}.

The PRISM algorithm introduced an additional concept to improve the scaling behaviour of STEM simulations via the scattering matrix. If only each $f$-th beam in the condenser aperture is sampled, the field of view in real-space contains $f^2$ copies of the probe with size ${\mathsf{N}_1/f^2 \cross \mathsf{N}_2/f^2}$. If one of these probe copies is cropped out, and then the far-field intensities computed via Eq.~\ref{equ:phase_shifting_multislice2}, we can perform simulations that trade a small amount of accuracy for a significant speed-up in computation times \cite{ophus2017fast}. The new model then reads 
\begin{equation}
    I(\vec{q},\mvec{\rho}) = \left|\f[\vec{r}]{\sum^{\mathsf{B}}_{\mathsf{b = 1}}\left[\mathbf{C}_{\mvec{\rho},\vec{r}}\,\mathcal{S}_{\vec{r},\mathsf{b}}\right]\Psi(\vec{h}_{\mathsf{b}})e^{-2 \pi i\vec{h}_{\mathsf{b}}\cdot\mvec{\rho}}}\right|^2, \nonumber
\end{equation}
where we have introduced a cropping operator,%
\begin{equation}
    \mathbf{C}_{\mvec{\rho},\vec{r}} = \begin{cases}
        1 & \mbox{if } \left|\vec{r} - \mvec{\rho}\right| \leq |\left|\mvec{\Delta}/2\right|\\
        0 & \mbox{otherwise}
 \nonumber
 \end{cases}
 \label{eq:prism}
\end{equation}
a two dimensional rectangular function of width ${\mvec{\Delta} \in \mathbb{R}^{\mathsf{N}_1 / f \cross \mathsf{N}_2 / f}}$ centered about each probe scan position $\mvec{\rho}$. This cropping procedure to compute STEM probes using PRISM is shown in Figs.~\ref{Figure:PRISM}c and d.

\section{The Partitioned PRISM Algorithm}

\subsection{Natural Neighbor Interpolation of the Scattering Matrix}

Theoretical and experimental investigations of \smatrix{} reconstructions have shown that once the plane wave tilts have been removed from all beams, the resulting matrix elements are remarkably smooth \cite{pelz2020reconstructing, brown2020three, findlay2021scatteringmatrix}. We have also observed that in many PRISM simulations, the information contained in neighboring beams is very similar. These observations have inspired us to propose a new method for simulating STEM experiments. Rather than computing all beams of the \smatrix{} with the multislice algorithm, we could instead interpolate them from a reduced set $\mathcal{P}$ of parent beams, which are computed with the multislice algorithm in the manner described above. 

Defining the interpolation weights as a matrix ${w \in \mathbb{R}^{|\mathcal{P}| x B}}$ that stores the interpolation weights for each beam, we can then compute the 4D-STEM intensities as
\begin{eqnarray}
    \label{equ:prism_beamlist}
     && I(\vec{q},\mvec{\rho}) = 
     \\
     && \left|\f[\vec{r}]{\sum^{\mathsf{B}}_{\mathsf{b = 1}}\Psi(\vec{h}_{\mathsf{b}})\,e^{-2 \pi i \vec{h}_{\mathsf{b}}\cdot\bm{\rho}}\,\mathbf{C}_{\mvec{\rho},\vec{r}}\left[\sum_{\mathsf{p} \in \mathcal{P}}w_{\mathsf{p},\mathsf{b}}\mathcal{S}_{\vec{r},\mathsf{b}}\right]}\right|^2. \nonumber 
\end{eqnarray}


The remaining tasks are then to choose an interpolation strategy to determine the weights $w$, and to choose the set of parent beams $\mathcal{P}$. To maximise the flexibility in choosing the parent beams, which form the the interpolation bases of the \smatrix{}, the chosen interpolation scheme must be able to interpolate an unstructured grid of parent beams. Here we have chosen to employ the natural neighbour interpolation \cite{sibson1981brief, amidror2002scattered}.

We note two additional methods which can save further computational time. First, part of the computational overhead when performing matrix multiplication of the \smatrix{} is the cropping operator. When using interpolation factors of $f>1$ for either traditional or partitioned PRISM, this overhead can represent a significant amount of computation time due to the need for a complex indexing system to reshape a subset of the \smatrix{}. Thus in many cases, simulations with $f=2$ may require longer computational times than $f=1$. We therefore recommend that the scaling behaviour be tested in each case.


\begin{figure*}[htbp]
    \centering
    \includegraphics[width=6.4in]{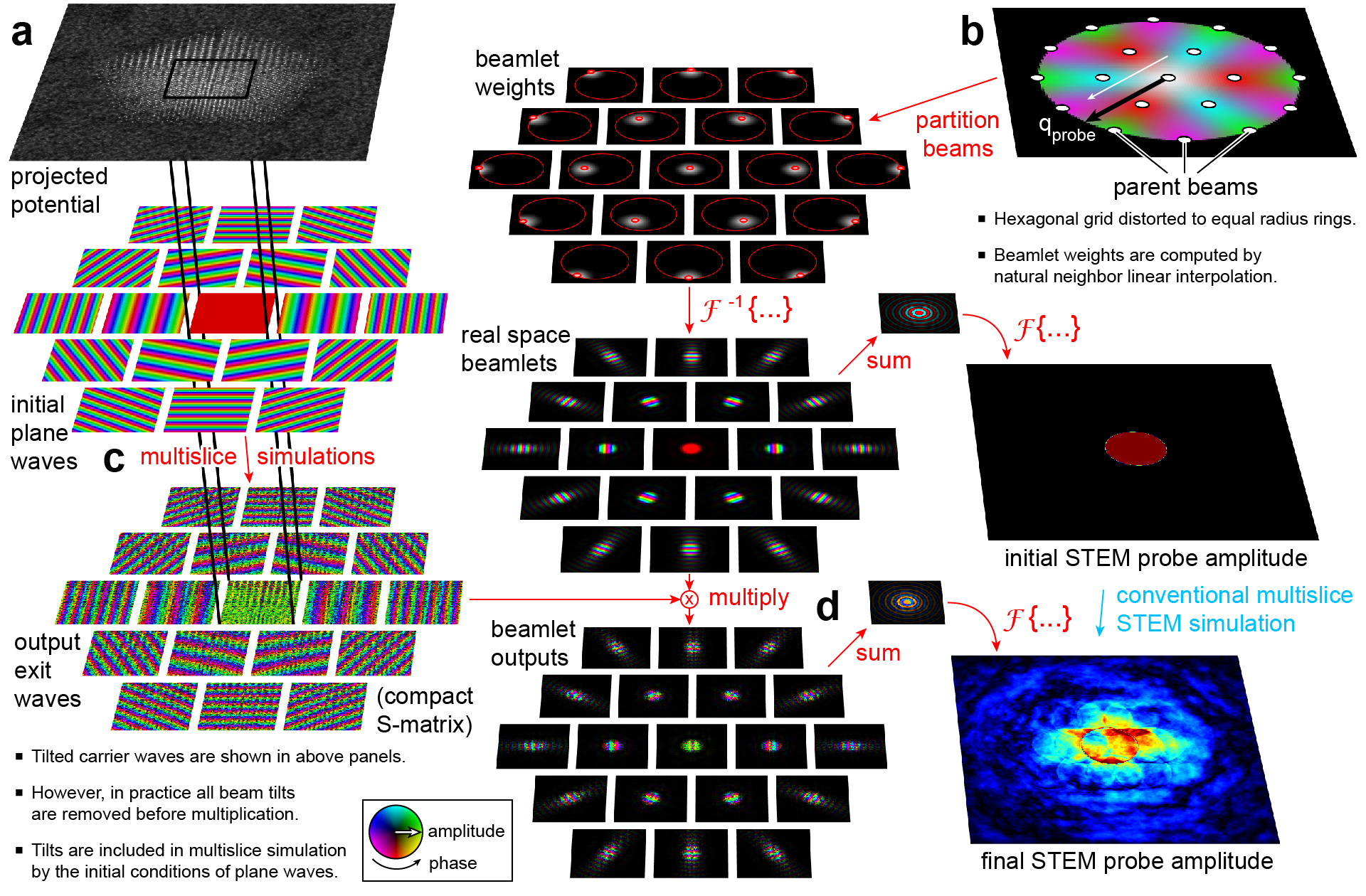}
    \caption{{\bf Flow chart of the beam partitioning algorithm for STEM simulation.} (a) Calculate the projected potential slices from the atomic coordinates and lookup tables. (b) Partition the probe into the desired number of beams, calculate the basis functions for all beamlets. (c) Perform a multislice simulation of all beams defined by the partitioning, store results in compact \smatrix{}. (d) Construct STEM probes at all positions by multiplying the shifted initial probes by the \smatrix{}, and then summing over all beams.}
    \label{Figure:partPRISM}
\end{figure*}

A second and more universal speed-up can be achieved by modifying the beamlet basis functions. To position a STEM probe at any position that is not exactly centered on a pixel with respect to the plane wave basis functions, we use the Fourier shift theorem to apply the sub-pixel shifts of the initial probe, represented by the phase factors $e^{-2 \pi i \vec{h}_{\mathsf{b}}\cdot\bm{\rho}}$ in Equ. \ref{equ:prism_beamlist}. This requires that the beamlet basis functions be stored in Fourier space coordinates, multiplied by a plane wave to perform the sub-pixel shift, and then an inverse Fourier transform be performed before multiplication by the \smatrix{}. To avoid this potentially computationally-costly step, we can set the simulation parameters such that all STEM probe positions fall exactly on the potential array pixels (for example, calculating 512 x 512 probe positions from a 1024 x 1024 pixel size potential array). This eliminates the summation over the complete set of basis beams $\mathsf{b}$ and the shift of the STEM probe can be achieved by indexing operations alone, allowing the probe basis functions to be stored in real space. 
After factoring out the summation over the Fourier basis, the 4D-STEM intensities can then be calculated as
\begin{equation}
     I(\vec{q},\mvec{\rho}) = \left|\f[\vec{r}]{\sum_{\mathsf{p} \in \mathcal{P}}\hat{\psi}_{\mathsf{p}}\,\mathbf{C}_{\mathsf{k},\mathsf{d}}\left[\mathcal{S}_{\vec{r},\mathsf{p}}e^{2 \pi i \vec{h}_{\mathsf{p}}\cdot\vec{r}}\right]}\right|^2,
\label{equ:prism_beamlist_pixel}
\end{equation}
with
\begin{equation}
    \hat{\psi}_{\mathsf{p}} = \sum^{\mathsf{B}}_{\mathsf{b = 1}} w_{\mathsf{p},\mathsf{b}} \Psi(\vec{h}_\mathsf{b}) e^{-2 \pi i \vec{h}_\mathsf{b}\cdot\vec{r}}
\end{equation}
which defines our new ``beamlet'' basis, depicted in Fig. \mbox{\ref{Figure:partPRISM} a)}. These new probe basis functions can be pre-computed and stored in memory, such that only summation over the the parent beams is necessary to calculate a diffraction pattern.

\subsection{Algorithmic Steps of Partitioned PRISM Simulations}

Fig.~\ref{Figure:partPRISM} shows the steps of our new simulation algorithm as a flow chart. As in all of the above electron scattering simulation methods, the first step is to compute the projected potentials from the atomic coordinates. Fig.~\ref{Figure:partPRISM}a shows the sum of all 40 projected potential slices, each having a thickness of 2 \AA.

The second step, shown in Fig.~\ref{Figure:partPRISM}b is to choose a set of parent beams, and then calculate the weight function of all beamlets using the desired partitioning scheme. Here we have used two rings of triangularly tiled beams, where each ring has a constant radius and the beams are separated by 10 mrads across the 20 mrad STEM probe. The beamlet weights $w$ are calculated using natural neighbor interpolation, and are shown in Fig.~\ref{Figure:partPRISM} a) in the top right panel. The parent beams are indicated by small red circles in the condenser aperture, and the beamlet weight distributions for each parent beam are show in gray scale. By taking the inverse Fourier transform of each weight function, we can generate the real space beamlet basis functions $\hat{\psi}_{\mathsf{p}}$.

Fig.~\ref{Figure:partPRISM} c) shows the third step of the partitioning simulation algorithm, where we perform a plane wave multislice simulation for each of the parent beams defined above. After the plane waves have been propagated and transmitted through all 40 slices, the tilt of each beam is removed. These outputs are then stored in the compact \smatrix{} $\mathcal{S}_{\vec{r},\mathsf{p}}$.

Finally, we compute the intensity of each desired STEM probe position as shown in Fig.~\ref{Figure:partPRISM} d). First, if we are using a PRISM interpolation factor other than $f=1$, we crop out a subset of the \smatrix{}. Next, each beam of the \smatrix{} is multiplied by the beamlet basis functions, and all beamlet exit waves summed to form the complex STEM probe in real space. Finally, we take the Fourier transform of these probes and compute the intensity from the magnitude squared of the wavefunction. If we require sub-pixel shifts of the STEM probes with the cropped region of the \smatrix{}, we must first multiply the probe basis functions by the appropriate complex plane wave in Fourier space to achieve the desired shift. This adds some computational overhead to each probe, and so if possible we suggest using a potential sampling pixel size that produces a simulation image size which is an integer multiple of the spacing between adjacent STEM probes.

\section{Computational and memory complexity}

We now approximate computational complexity and memory complexity for the multislice, PRISM, and partitioned PRISM algorithms. We neglect calculation time for the sample projected potential slice and thermal diffuse scattering, as the added computational and memory complexity is equal for all methods. For simplicity, we assume a quadratic simulation cell with $\mathsf{N} = \mathsf{N}_1 = \mathsf{N}_2$. Each slice of the multi-slice algorithm requires transmission and propagation operations in Equ. \ref{eq:multiply_propagate}, which is $6\mathsf{N}^2 \log_2(\mathsf{N})$, and $2\mathsf{N}^2$ operations to multiply the potential and the Fresnel propagator. For a STEM simulation with $\mathsf{P}$ STEM probe positions and $\mathsf{H}$ slices for the sample, the total multi-slice complexity is then $\Theta(\mathsf{H}\mathsf{P}(6\mathsf{N}^2 \log_2(\mathsf{N})+2\mathsf{N}^2)$ \cite{ophus2017fast}. The complexity of the PRISM algorithm is given by $\Theta(\frac{\mathsf{H}\mathsf{B}}{f^2}\left[6\mathsf{N}^2 \log_2(\mathsf{N})+2\mathsf{N}^2\right]+\frac{\mathsf{P}\mathsf{B}\mathsf{N}^2}{4f^4})$ \cite{ophus2017fast}, which consists of $\mathsf{H}\mathsf{B}$ multi-slice simulations for each of the sampled beams, and $\frac{\mathsf{P}\mathsf{B}\mathsf{N}^2}{4f^4}$ operations for the summation of the beams. For the partitioned PRISM algorithm with $\mathsf{B}_p$ partitions, the complexity for the multi-slice calculations is $\Theta(\frac{H\mathsf{B}_p}{f^2}\left[6\mathsf{N}^2 \log_2(\mathsf{N})+2\mathsf{N}^2\right])$. For the real-space summation with subpixel precision, a maximum of $\frac{\mathsf{P} \mathsf{B} \mathsf{B}_p \mathsf{N}^2}{4f^2}$ operations is necessary, while for the integer positions on the \smatrix{}-grid, only $\frac{\mathsf{P}\mathsf{B}_p \mathsf{N}^2}{4f^2}$ operations are necessary.

\begin{table*}[htbp]
  \begin{center}
  \small
  \begin{tabularx}{\textwidth}{Xp{6cm}p{3.5cm}}
    {\bf Algorithm} & 
    {\bf Time Complexity} & 
    {\bf Memory Complexity} \\
    \hline
    Multislice & $\Theta(\mathsf{H}\mathsf{P}(6\mathsf{N}^2 \log_2(\mathsf{N})+2\mathsf{N}^2)$& $\Theta(\mathsf{N}^2+\frac{\mathsf{P}\mathsf{N}^2}{f^2})$ \\
    PRISM &  $\Theta(\frac{\mathsf{H}\mathsf{B}}{f^2}\left[6\mathsf{N}^2 \log_2(\mathsf{N})+2\mathsf{N}^2\right]+\frac{\mathsf{P}\mathsf{B}\mathsf{N}^2}{4f^4})$ & $\Theta(\frac{\mathsf{B}\mathsf{N}^2}{4}+\frac{\mathsf{P}\mathsf{N}^2}{f^2})$ \\
    partitioned PRISM subpixel precision & $\Theta(\frac{\mathsf{H}\mathsf{B}_p}{f^2}\left[6\mathsf{N}^2 \log_2(\mathsf{N})+2\mathsf{N}^2\right]+\frac{\mathsf{P} \mathsf{B} \mathsf{B}_p \mathsf{N}^2}{4f^2})$ & $\Theta(\frac{\mathsf{B}_p \mathsf{N}^2}{4}+\frac{\mathsf{P}\mathsf{N}^2}{f^2})$ \\
    partitioned PRISM integer pixel precision& $\Theta(\frac{\mathsf{H}\mathsf{B}_p}{f^2}\left[6\mathsf{N}^2 \log_2(\mathsf{N})+2\mathsf{N}^2\right]+\frac{\mathsf{P}\mathsf{B}_p \mathsf{N}^2}{4f^2})$ & $\Theta(\frac{\mathsf{B}_p\mathsf{N}^2}{4}+\frac{\mathsf{P}\mathsf{N}^2}{f^2})$ \\
  \end{tabularx}
  \caption{
  \textbf{Computational and memory complexity of alternatives for computing scanning transmission electron microscopy data.}
  \textcolor{darkred}{$\mathsf{H}$: number of slices, $\mathsf{B}$: number of beams, $\mathsf{B}_p$: number of beam partitions, $\mathsf{P}$: number of probes, $\mathsf{N}$: side length of field of view in pixels.}
  }
  \label{tab:interactions_alternatives}
  \vspace{-0.2cm}
  \end{center}
\end{table*}

The memory complexity of the multi-slice algorithm is lowest, since only the current wave of size $\Theta(\mathsf{N}^2)$ needs to be held in memory for an unparallelized implementation. All algorithms need $\Theta(\frac{\mathsf{P}\mathsf{N}^2}{f^2})$ memory to store the results of the calculation, if 4D datasets are computed. The PRISM algorithm requires $\Theta(\frac{\mathsf{B}\mathsf{N}^2}{4})$ memory to store the compact \smatrix{}. For simulations which require a finely-sampled diffraction disk, $\mathsf{B}$ can quickly grow to $10^4$ or larger, since the number of beams scales with the square of the bright-field disk radius, such that large-scale simulations with fine diffraction disks can outgrow the available memory on many devices. The memory requirements of the partitioned PRISM algorithm scale with $\Theta(\frac{\mathsf{B}_p \mathsf{N}^2}{4})$. Since the number of parent beams $\mathsf{B}_p$ can be chosen freely, the memory requirements of the partitioned PRISM algorithm can be freely adjusted to the available hardware.

\section{Methods}

All the simulations shown in this paper were performed using a set of custom Matlab codes. In addition to implementing the partitioned PRISM algorithm, we have also implemented both the conventional multislice and PRISM algorithms for STEM simulation, in order to make a fair comparison between the different methods. We have used a single frozen phonon configuration in all cases, in order to increase the number of features visible in diffraction space. No effort was made for performance optimization or parallelization beyond MATLAB's inline compiler optimizations.   

The microscope parameters used in Figs~\ref{Figure:singleProbesPartition},\ref{Figure:singleProbesPartitionPRISM} and \ref{Figure:STEMimages} were an accelerating voltage of \SI{80}{\kilo\volt}, a probe convergence semiangle of \SI{20}{\milli\radian}, and a pixel size of \SI{0.1}{\angstroms}. The probe was set to zero defocus at the entrance surface of the simulation cell. The projected potentials were calculated using a 3D lookup table method \cite{rangeldacosta2021prismatic2p0}, using the parameterized atomic potentials given in \cite{kirkland2020}. Slice thicknesses of 2 \AA~were used for all simulations, and an anti-aliasing aperture was used to zero the pixel intensities at spatial frequencies above $0.5 \cdot q_\text{max}$ during the propagation step.

The atomic coordinates utilized for our single probe position and imaging simulations is identical to that used previously \cite{ophus2017fast}. The structure consists of a Pt nanoparticle with a multiply twinned decahedral structure, with screw and edge dislocations present in two of the grains. The nanoparticle measures approximately 7 nm in diameter, and was tilted such that two of the platinum grains are aligned to a low index zone axis. It was embedded into an amorphous carbon support to a depth of approximately 1 nm, with all overlapping carbon atoms removed. The cell size is \SI{10 x 10 x 8}{\nano\meter}, and contains \num{57443} total atoms. The nanoparticle coordinates were taken from \cite{chen2013three}, and the amorphous carbon structure was adapted from \cite{ricolleau2013random}. 


The atomic coordinates of our 4D-STEM simulations were a multilayer stack of semiconductor materials inspired by the experiments in \cite{ozdol2015strain}. The simulation cell consists of a GaAs substrate where the Ga and As sites are randomly replaced with 10\% Al and P respectively. The multilayers are an alternating stack of GaAs doped with 10\% P and pure GaAs respectively, each 9 unit cells thick along a [001] direction. The lattice parameters of the GaAs and GaAsP were fixed to be +1.5\% and -1.5\% of the substrate lattice parameter, which was set to 5.569 \AA. The field of view was approximately 500 x 500 \AA, and the potential pixel size and slice thicknesses were set to \SI{0.1}{\angstroms} and \SI{2}{\angstroms} respectively. The cell thickness was approximately \SI{40}{\angstroms}, giving 4.6 million atoms inside the simulated volume. The STEM probe convergence semiangle was set to 2.2 mrads, the accelerating voltage was set to \SI{300}{\kilo\volt}, and the probe was scanned over the field of view with \SI{2}{\angstroms} step sizes, giving an output of \num{250 x 250} probes. 

The simulations shown in Figs.~\ref{Figure:singleProbesPartition}, \ref{Figure:singleProbesPartitionPRISM}, and Fig.~\ref{Figure:sim4DSTEM} were computed on a laptop with an Intel Core i7-10875H GPU, operating at 2.30 GHz with 8 cores, and 64 GB of DDR4 RAM operating at 2933MHz. The simulations shown in Fig. \ref{Figure:STEMimages} were performed on Intel Xeon Processors E5-2698v3 with 8 Physical cores (16 threads) and 25GB RAM per simulation. The multislice \num{512x512} results were obtained by splitting the \num{512x512} array into 32 jobs with \num{16x512} positions. Prism $f=2$ results with \num{512x512} probe positions were obtained by splitting the array in to \num{8} jobs with \num{64x512} each, all using an identical calculated \smatrix{}. All calculations were performed using Matlab's single floating point complex numbers, and simulation run times were estimated using built-in MATLAB functions, and memory usages were based on theoretical calculations. 


\section{Results and Discussion}

\subsection{Calculation of Individual STEM Probes}

\begin{figure}[htbp]
    \centering
    \includegraphics[width=3.1in]{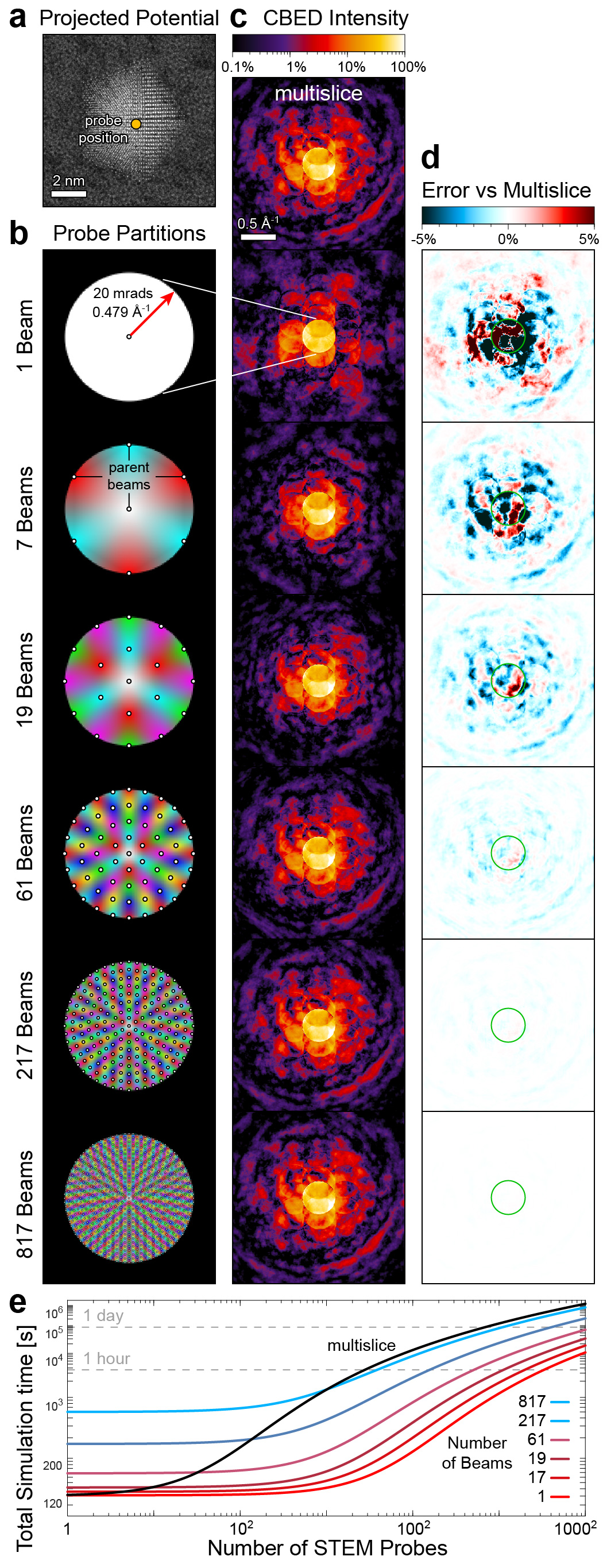}
    \caption{{\bf Individual STEM probes computed with beam partitioning.} (a) Projected potential and probe position. (b) Partitioning diagram showing beamlet weights. (c) Calculated CBED intensity on a logarithmic scale. (d) Error versus multislice probe simulation. (e) Estimated calculation time of the different probe partitions as a function of the number of probes calculated.}
    \label{Figure:singleProbesPartition}
\end{figure}

To demonstrate the accuracy of our proposed algorithm, we have performed STEM simulations of a common sample geometry: a multiply-twinned Pt nanoparticle resting on an amorphous surface. The total projected potential of this sample is plotted in Fig.~\ref{Figure:singleProbesPartition}a, as well as the location of a STEM probe positioned just off-center. We have tested a series of beam partitioning schemes, shown graphically in Fig.~\ref{Figure:singleProbesPartition}b. The first case tested was a single beam, which is equivalent to convolving a plane wave HRTEM simulation with the STEM probe. We have also used natural neighbor partitioning to calculate the beamlet weights when using a series of concentric hexagonal rings of beams, distorted slightly to the circular probe geometry. These simulations include partitioning the 20 mrad  probe by 20, 10, 5, 2.5, and 1.25 mrads, resulting in a total of 7, 19, 61, 217 and 817 parent beams respectively. 

The calculated diffraction space intensities of the probes corresponding to the above cases are shown in Fig.~\ref{Figure:singleProbesPartition}c, along with the corresponding conventional multislice simulation. We see that using a single parent beam is extremely inaccurate, reproducing only the coarsest features of the multislice simulation. However, the partitioning scheme rapidly converges to the multislice result, shown by the error images plotted in Fig.~\ref{Figure:singleProbesPartition}d. The 19 beam case has errors falling roughly within 5\%, while the 61 beam case drops to $<$2\%. The calculated probe for the 217 beam case has errors on the order of  $<$0.5\%, which would likely be indistinguishable from an identical experiment due to measurement noise. Finally, the 817 beam case is essentially error-free.

We can make additional observations about the character of the errors present in the partitioning algorithm. Inside the initial probe disk and in directly adjacent regions, the errors are roughly equally distributed in the positive and negative directions. However, at higher angles the errors are biased in the negative direction. This indicates that the partitioning approximation is highly accurate at low scattering angles where coherent diffraction dominates the signal \cite{winkler2020direct}, and is less accurate at high scattering angles where thermal diffuse scattering dominates \cite{wang1989simulating}. We attribute this effect to the complex phase distribution of the pixels; at low scattering angles, adjacent beams have very similar phase distributions, which in turn makes the interpolation a good approximation. However, at high scattering angles the phases of each pixel are substantially more random, due to thermal motion of the atoms. This means that if too few beams are used to approximate the signal, the coherent summations will tend towards zero due to the random phase factors. Thus when using a small number of beams in partitioned STEM simulations, high angle scattering intensities can be underestimated.

The estimated calculation times for these simulations are shown in Fig.~\ref{Figure:singleProbesPartition}e. When calculating a single STEM probe, multislice is always fastest because the only overhead to the calculation is computation of the projected potentials. The partitioned simulations by contrast require evaluation of the \smatrix{}, which requires the same calculation time as each STEM probe multislice propagation for each beam. However, once the \smatrix{} has been computed, calculation of STEM probes via matrix multiplication becomes substantially faster than multislice. The overall simulation time becomes lower than multislice if many STEM probe positions must be calculated. For the 61, 217 and 817 beam cases, these crossovers in calculation time occur for 32, 135, and 1000 probe positions respectively. Therefore even for 1D simulations of STEM probe positions, the partitioning scheme is faster, and for simulations with a grid of 2D probe position this scheme is significantly faster than multislice.

\begin{figure}[htbp]
    \centering
    \includegraphics[width=3.1in]{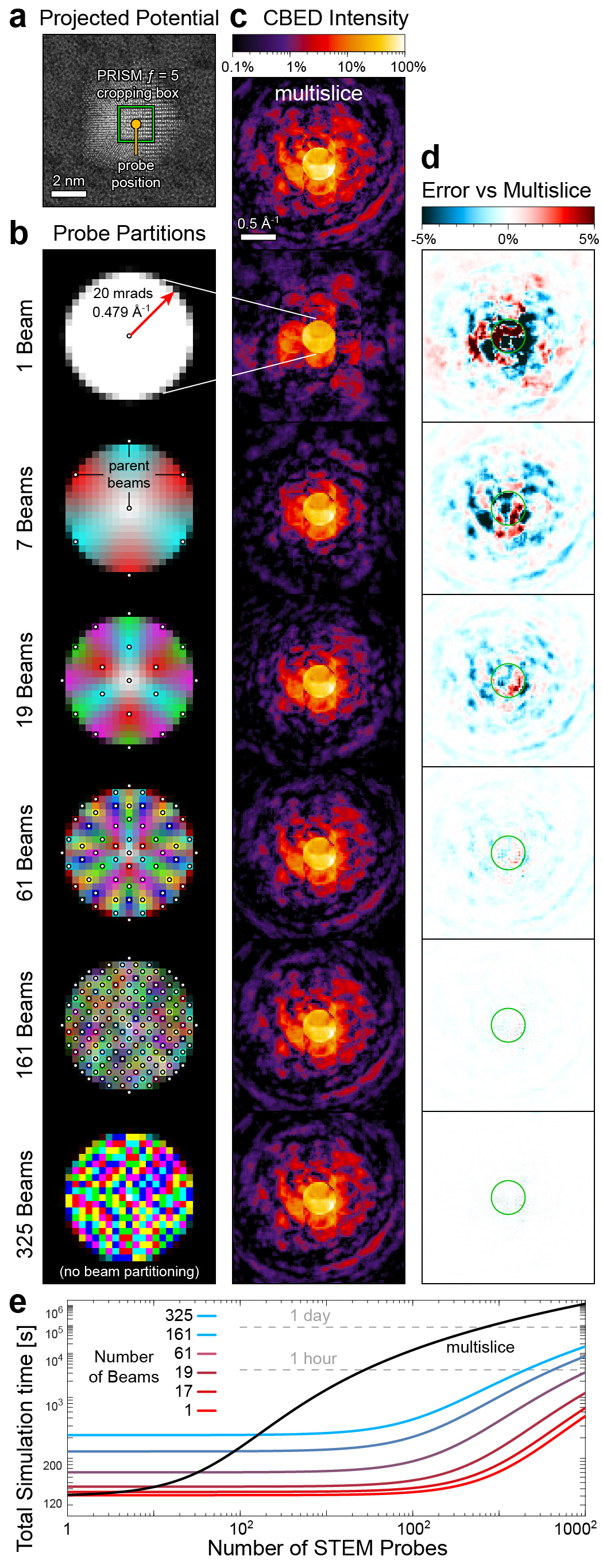}
    \caption{{\bf Individual STEM probes computed with beam partitioning and PRISM $f=5$ interpolation.} (a) Projected potential and probe position. (b) Partitioning diagram showing beamlet weights. (c) Calculated CBED intensity on a logarithmic scale. (d) Error versus multislice probe simulation. (e) Estimated calculation time of the different probe partitions as a function of the number of probes calculated.}
    \label{Figure:singleProbesPartitionPRISM}
\end{figure}

However, using the beam partitioning algorithm on the entire field of view does not utilize the algorithm to its full speed-up potential. The beam partitioning approximation is also compatible with the PRISM approximation. Partitioning reduces the number of entries of the \smatrix{} in diffraction space, whereas PRISM reduces the number of entries using cropping in real space. Fig.~\ref{Figure:singleProbesPartitionPRISM}a shows STEM simulations that combine partitioning with a PRISM interpolation factor of $f = 5$. The 25-fold reduction in sampling of the STEM probes is evident in Fig.~\ref{Figure:singleProbesPartitionPRISM}b, where the underlying beam pixels are clearly visible in the STEM probe. The partitioning scheme used is identical to that of Fig.~\ref{Figure:singleProbesPartitionPRISM}b, except for the 1.25 and 2.5 mrad partitioning cases. For the 1.25 mrad partitioning, the number of parent beams outnumbers the number of available beams; after removing duplicate beams, this simulation becomes equivalent to a PRISM $f=5$ simulation. The 2.5 mrad partitions were changed to a diagonal grid, where every other beam is included in order to produce a more uniform sampling of the \smatrix{}.

The calculated probe intensities are shown in  Fig.~\ref{Figure:singleProbesPartitionPRISM}c, along with the corresponding multislice simulation (which was sampled on the same 25-fold reduced grid). The errors of the partitioned PRISM simulations have been compared to the multislice simulation in  Fig.~\ref{Figure:singleProbesPartitionPRISM}d. The resulting convergence towards zero error is essentially identical to the non-PRISM case (where $f=1$). These simulations are also slightly biased towards negative errors at high scattering angles.

The estimated calculation times are plotted in Fig.~\ref{Figure:singleProbesPartitionPRISM}e, as a function of the number of probe positions. These simulations are substantially faster than multislice. The 61, 161 and 325 beam cases have a crossover in the calculation with multislice for 32, 83 and 155 probe positions respectively. If the error for the 161 beam case is within an acceptable tolerance, a 1000 x 1000 probe position simulation of this sample can be performed in roughly 50 minutes, without additional parallelization or utilization of GPU or HPC resources.

\subsection{Calculation of Full STEM Images}

\begin{figure*}[htbp]
    \centering
    \includegraphics[width=6.7in]{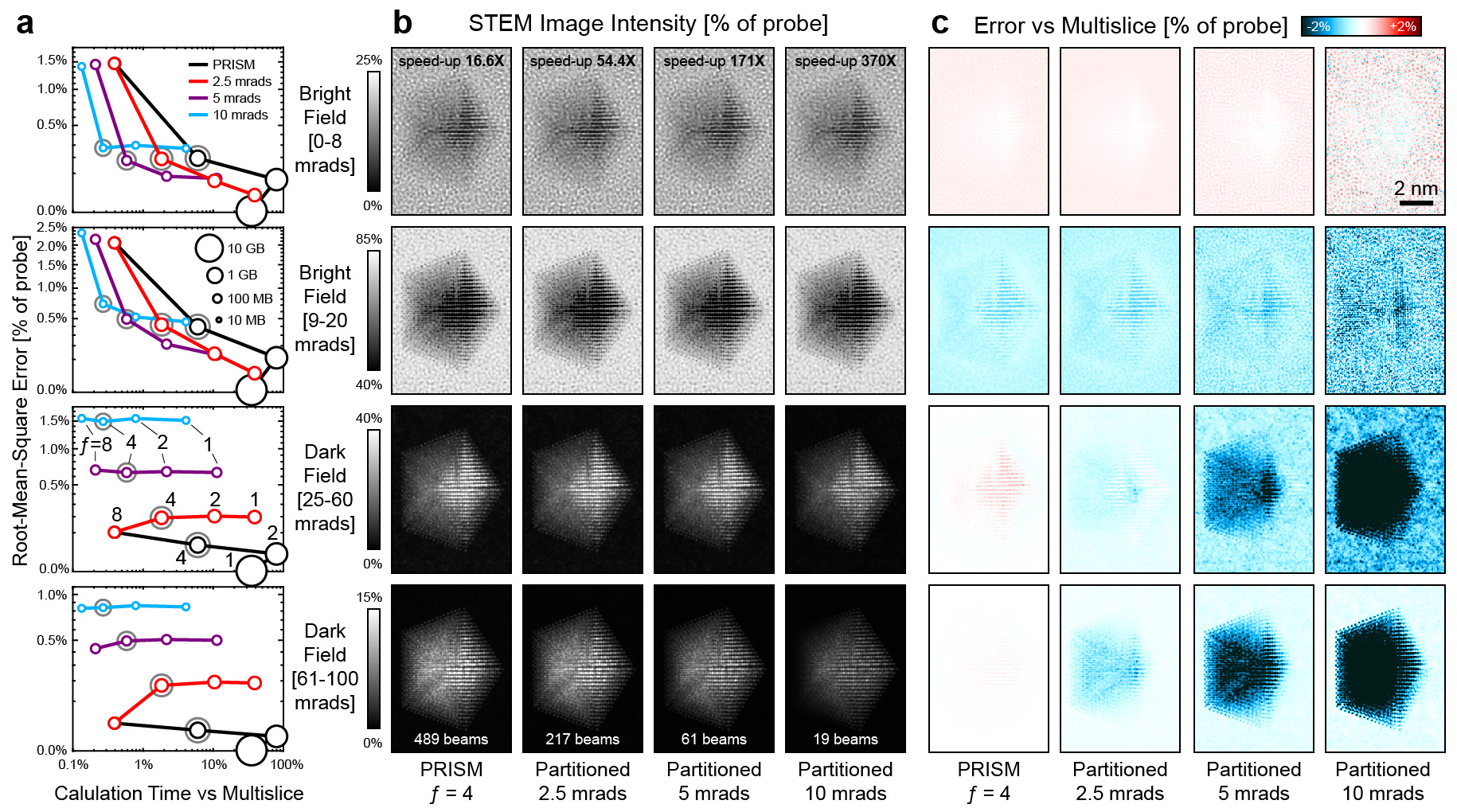}
    \caption{{\bf Simulation of STEM images using beam partitioning and PRISM interpolation.} (a) Calculation time, RMS error relative to multislice, and total RAM required for the \smatrix{} of all simulations. The four detector configurations considered are BF detector from 0-8 mrads, an ABF detector from 9-20 mrads, a LAADF detector from 25-60 mrads, and a HAADF detector from 61-100 mrads). Grey circles indicate the simulations where images are shown. (b) STEM images for the four detector configurations and four simulation cases labeled below. Calculation time speed-up relative multislice is inset into the top row, while the number of included beams is inset into the bottom row. (c) Pixel-wise errors of each image in (b) with respect to multislice simulations. }
    \label{Figure:STEMimages}
\end{figure*}

We have also simulated full STEM images with a variety of standard detector configurations, in order to demonstrate the potential of the partitioned PRISM algorithm. These simulations are shown in Fig.~\ref{Figure:STEMimages}, and include four radially-symmetric detector configurations. These are a bright field (BF) image from 0-8 mrads, an annular bright field (ABF) image from 9-20 mrads, a low angle annular dark field (LAADF) image from 25-60 mrads, and a high angle annular dark field (HAADF) image from 60-100 mrads. We have performed these simulations with 512 x 512 probe positions using  multislice, PRISM and partitioned PRISM algorithms. The PRISM simulations used interpolation factors of $f=1$, 2, 4, and 8, giving a total number of beams equal to 7377, 1885, 489, and 137 beams respectively. The partitioning included was the scheme described above, where the 20 mrad STEM probe was subdivided by 10, 5, and 2.5 mrads into the parent beams, and where no partitioning was performed (i.e.\ the original PRISM algorithm). The number of beams for the 10, 5, and 2.5 mrad partitioning were equal to 19, 61, and 217 respectively, except for the 2.5 mrad partitioning for $f=8$ interpolation, where the simulation is equivalent to PRISM (137 beams).

Fig.~\ref{Figure:STEMimages}a shows a summary of the results, where the root-mean-square (RMS) errors in units of probe intensity and calculation times relative to multislice simulations are plotted. Additionally, the RAM requirements for storing the \smatrix{} are shown by the marker sizes. Overall, the results follow the same trend as in the previous section. Using less beams either in the partitioning or higher PRISM interpolation results in a less accurate simulation for all cases. The only exception to this is PRISM $f=1$ simulations, which are mathematically identical to multislice simulations \cite{ophus2017fast}. Interestingly, the PRISM ${f=1}$ simulations are faster than $f=2$, due to not needing any matrix indexing operations to crop out a portion of the \smatrix{}. However $f=1$ PRISM simulations also have the largest RAM requirements by a large margin, requiring 15.5 GB. This is potentially an issue for large simulations if we wish to utilize GPU resources, since RAM capacities of current GPUs are often in the range of 4-16 GB, and there is additional overhead for other arrays that must be calculated. This problem can be alleviated by streaming only part of the \smatrix{} into the GPU RAM \cite{pryor2017streaming}, but then the large speed-up afforded by performing only a single matrix multiplication per STEM probe is lost.

The BF and ABF simulation errors shown in Fig.~\ref{Figure:STEMimages}a, the partitioned simulations have a very favourable balance between calculation time and accuracy. For PRISM interpolation factors of $f=2$ and $f=4$, the partitioned simulations have essentially identical accuracy to the PRISM simulations, while requiring far lower calculation times and less RAM to store the \smatrix{}. The 5 mrad partitioning case (61 beams) for example is 46 ($f=2$) and 171 ($f=4$) times faster than an equivalent multislice simulation, while having RMS errors on the order of 0.2\% and 0.1\% respectively for the 0-8 mrads BF image and RMS errors on order of 0.5\% and 0.2\% respectively for the 9-20 mrads ABF image. 

For the LAADF and HAADF images shown in Fig.~\ref{Figure:STEMimages}a, the partitioned simulations show somewhat less favourable error scaling than the PRISM algorithm. While the calculation times are reduced by partitioning for a given PRISM interpolation factor, the errors increase roughly inversely proportional to the number of include beams. These errors are still relatively low however, staying roughly constant with the interpolation factor $f$.

Fig.~\ref{Figure:STEMimages}b shows the STEM images for the $f=4$ cases including conventional PRISM and the 3 partitioning schemes. It is immediately evident that all images contain the same qualitative information, for example showing that the ABF image is far more interpretable than the BF image. Visually, the BF and ABF images appear indistinguishable from each other, with all atomic-scale features preserved across the different partitioning schemes. The LAADF and HAADF images similarly all contain the same qualitative information, and all highlight the differences between these two dark field imaging conditions. Here however we can see an overall reduction of image intensity inside the nanoparticle for the partitioned simulations with less beams. In the LAADF case, the 19 beam image is noticeably dimmer than the other cases, and for the HAADF case both the 19 and 61 beam partitioning show reduced intensities.  

To show the errors more quantitatively as a function of the probe position, we have plotted the difference images with respect to a multislice simulation in Fig.~\ref{Figure:STEMimages}c. For the BF images, a slight offset in the overall intensities is visible, likely due to the slightly different probe and detector sampling when using $f=4$ interpolation. The spatially resolved differences are very low however, for both PRISM and the 2.5 and 5 mrad partitioning simulations. In the regions of highest scattering in the nanoparticle, some errors along the atomic planes are visible in the 10 mrad partitioned simulation. The 2.5 and 5 mrads partioned PRISM simulations are an excellent replacement for the PRISM simulations, as they offer large calculation time speed-ups for a negligible change in the error.

In the LAADF and HAADF error images plotted in Fig.~\ref{Figure:STEMimages}c, the errors are increasing proportionally to the inverse of the number of beams included, as we observed in Fig.~\ref{Figure:STEMimages}a. The HAADF images show higher overall errors than the LAADF images, due to the increasing randomness of the pixel phases at high scattering angles where thermal diffuse scattering dominates the signal. The relative error is also higher at these high scattering angles, as the number of electrons that scatter to these high angles are significantly lower than those which reach the other detector configurations. Both the LAADF and HAADF errors scale nearly linearly with the nanoparticle projected potential, which indicate that they may be tolerable for relative measurements such as comparing different thicknesses or the signal measured for different atomic species. For quantitative intensity simulations at high angles, we recommend using as many beams as possible for the partitioned PRISM algorithm.


\subsection{Calculation of 4D-STEM Datasets}

\begin{figure}[htbp]
    \centering
    \includegraphics[width=3.3in]{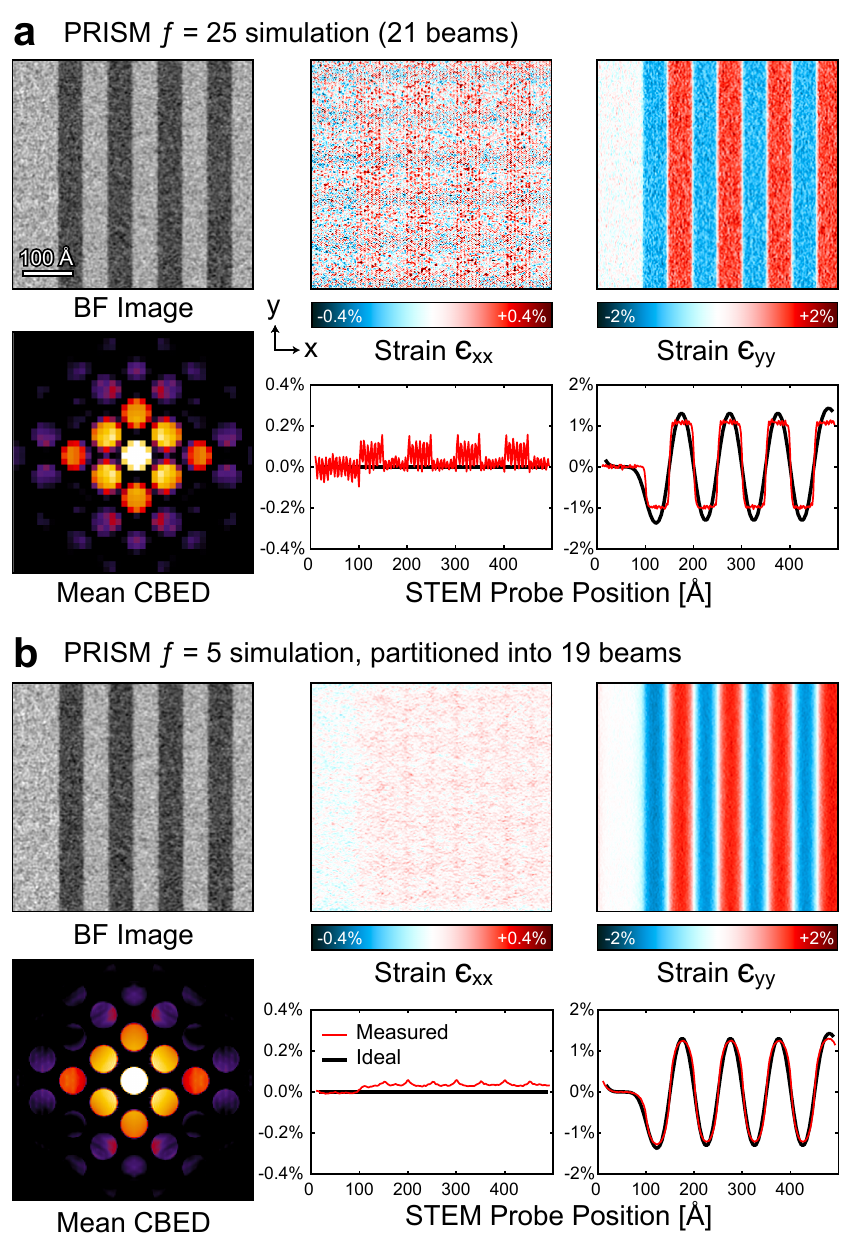}
    \caption{{\bf Simulated 4D-STEM datasets and strain maps of a multilayer semiconductor stack.} (a) PRISM simulation with $f=25$ interpolation and 21 total beams. (b) Partitioned PRISM simulation with $f=5$ interpolation and 19 total beams. Each simulation shows a virtual bright field image, the mean CBED image, and strain maps in the two cardinal directions. Line traces show average strain perpendicular to the layer direction.}
    \label{Figure:sim4DSTEM}
\end{figure}


Many 4D-STEM experimental methods require fine enough sampling of reciprocal space to resolve the edges of scattered Bragg disks, or fine details inside the unscattered and scattered Bragg disks \cite{ophus2019four}. In particular, for machine learning methods which are trained on simulated data, we want the sampling and image sizes to be as close to the experimental parameters as possible \cite{xu2018deep, yuan2021training}. Here, compare the PRISM and partitioned PRISM algorithms for 4D-STEM simulations, and assess their accuracy by performing a common 4D-STEM workflow of strain mapping by measuring the Bragg disk spacing \cite{pekin2017optimizing}.

We have simulated a 4D-STEM experiment for a multilayer stack of semiconductor materials similar to the experiments shown in \cite{ozdol2015strain}, shown in Fig.~\ref{Figure:sim4DSTEM} and described above. Two simulations were performed: the first used only the PRISM algorithm with interpolation factors of $f=25$, giving 21 total beams, shown in Fig.~\ref{Figure:sim4DSTEM}a. The second combined a PRISM interpolation of $f=5$ with partitioning into 19 beams, shown in Fig.~\ref{Figure:sim4DSTEM}b. These parameters were chosen to require approximately the same total calculation time (157 and 186 minutes for pure PRISM and partitioned PRISM respectively). Both simulations used the same atomic potentials which required 113 minutes to compute. The \smatrix{} calculation steps required 42 and 30 minutes for the pure PRISM and partitioned PRISM simulations respectively. Finally the \num{62500} probe positions required 2 and 43 minutes for the pure PRISM and partitioned PRISM simulations respectively.

We have used the py4DSTEM package \cite{savitzky2020py4dstem} to measure strain in both of the simulations shown in Fig.~\ref{Figure:sim4DSTEM}, by fitting the positions of the Bragg disks. These strains are compared to the ideal strain, estimated by convolving the underlying lattice spacing with a Gaussian kernel with a standard deviation given by the \SI{5}{\angstroms} estimated size of the STEM probe. In the pure PRISM simulation shown in Fig.~\ref{Figure:sim4DSTEM}a, there are artifacts visible in both strain maps. The strain perpendicular to the layer direction shows rapid oscillations of $\pm0.1\%$, while the strain parallel to the layer direction shows discrete steps. Both of these are due to the very small cropping box used when $f=25$, which cuts off the tails of the STEM probe in this simulation. Additionally, the limited sampling of the diffraction disk edges strongly limits the achievable precision in the disk position measurements.

By contrast, the partitioned simulation shown in Fig.~\ref{Figure:sim4DSTEM}b samples diffraction space 5 times more finely in both the $x$ and $y$ directions. The resulting strain maps are much flatter, and the measured strain positions agree better with the ideal measurements. This simulation demonstrates that beam partitioning combined with PRISM interpolation can provide a much more efficient use of the calculation time required to generate the \smatrix{} beams than a pure PRISM simulation. This partitioning case uses approximately the same number of beams and requires roughly the same calculation time, but is substantially more accurate at the low scattering angles used in a coherent diffraction 4D-STEM simulation. We also estimate that a multislice simulation of this same experiment would require approximately 60 days using the same simulation parameters. Even if we were to increase the beam sampling by a factor of 8, the partitioned PRISM simulation would still complete in less than a day.



\section{Conclusion}

We have introduced the beam partitioning algorithm for STEM simulation. This algorithm splits the STEM probe into a series of basis functions generated by natural neighbor interpolation between a set of parent beams. We construct the diffracted STEM probe by matrix multiplication of these basis functions with plane wave multislice simulations of each parent beam which are stored in a \smatrix{} that can be re-used for each new STEM probe position. We have demonstrated that the resulting algorithm converges rapidly to low error with respect to the conventional multislice algorithm, and that it is fully compatible with the PRISM algorithm for STEM simulation. 

We have compared our new algorithm to multislice and PRISM simulations of a nanoparticle on an amorphous substrate.  With these simulations, we have shown that in general partitioned beam simulations can provide the same accuracy as PRISM at low to intermediate scattering angles (where coherent diffraction dominates the signal), but with much lower calculation times and lower RAM usage. We have also shown that at high scattering angles, beam partitioning simulation accuracy is somewhat worse than the PRISM algorithm, though still with lower calculation times. These low calculation times may allow the partitioned PRISM algorithm to be used ``in the loop'' with 3D tomographic reconstruction algorithms, in order to properly model the nonlinear dependence of STEM image contrast on the underlying atomic potentials.

Finally, we have also demonstrated the utility of partitioned PRISM for simulations of large 4D-STEM datasets. We used a common sample geometry composed of a multilayer stack of semiconductor materials with varying compositions on a substrate, and performed strain mapping from the diffracted probe signals by measuring the position of the Bragg disks and fitting a lattice. These simulations show that the partitioned PRISM algorithm is particularly well suited for performing fast simulations of large fields of view where high sampling of diffraction space is needed. We believe that our algorithm will find widespread application in simulations of very large simulated cells, such as those calculated with molecular dynamics. Our simulations  also show that the beam partitioning \smatrix{} can efficiently represent complex three-dimensional scattering, which may make it useful for inverting experimental data efficiently. 

\section{Data and Source Code Availability}

Reference implementations of the algorithms presented in this paper (multislice, PRISM and partitioned beam STEM simulations) are available at \href{https://github.com/cophus/superPRISM}{github.com/cophus/superPRISM}.

\section*{Acknowledgements}

PP and MCS are supported by the Strobe STC research center. LRD and BHS are supported by the Toyota Research Institute; LRD is also supported by the Department of Energy Computational Science Graduate Fellowship. CO acknowledges support from the Department of Energy Early Career Research Award program.  Work at the Molecular Foundry was supported by the Office of Science, Office of Basic Energy Sciences, of the U.S. Department of Energy under Contract No. DE-AC02-05CH11231. This material is based upon work supported by the U.S. Department of Energy, Office of Science, Office of Advanced Scientific Computing Research, Department of Energy Computational Science Graduate Fellowship under Award Number DE-SC0021110.

\bibliographystyle{naturemag}
\bibliography{refs}    

\end{document}